\documentclass[aps,prb,twocolumn,superscriptaddress,showpacs]{revtex4}
\usepackage{epsfig}
\usepackage{graphicx}
\usepackage{amsfonts}
\usepackage[figuresright]{rotating}
\usepackage{amssymb}
\usepackage{amsmath}
\usepackage{psfrag}
\def\avg#1{\langle#1\rangle}

\def\be{\begin{equation}}
\def\ee{\end{equation}}
\def\bea{\begin{eqnarray}}
\def\eea{\end{eqnarray}}

\def\nn{\nonumber}

\def\sgn{\mbox{sgn}}

\begin{document}
\title{2D and 3D topological insulators with isotropic and parity-breaking
Landau levels}
\author{Yi Li}
\affiliation{Department of Physics, University of California, San Diego,
La Jolla, California 92093, USA}
\author{Xiangfa Zhou}
\affiliation{
Key Laboratory of Quantum Information, University of Science and
Technology of China, CAS, Hefei, Anhui 230026, China}
\author{Congjun Wu}
\affiliation{Department of Physics, University of California,
San Diego, CA 92093}

\begin{abstract}
We investigate topological insulating states in both two and three dimensions
with the harmonic potential and strong spin-orbit couplings breaking the
inversion symmetry.
Landau-level like quantization appear with the full two- and three-dimensional rotational
symmetry and time-reversal symmetry.
Inside each band, states are labeled by their angular momenta over which
energy dispersions are strongly suppressed by spin-orbit coupling to being nearly flat.
The radial quantization generates energy gaps between neighboring bands at the
order of the harmonic frequency.
Helical edge or surface states appear on open boundaries characterized
by the Z$_2$ index.
These Hamiltonians can be viewed from the dimensional reduction of the
high dimensional quantum Hall states in 3D and 4D flat spaces.
These states can be realized with ultra-cold fermions inside harmonic
traps with the synthetic gauge fields.
\end{abstract}
\maketitle

\section{Introduction}
\label{sect:intro}

The study of topological insulators has become an important
research focus in condensed matter physics \cite{hasan2010,qi2011}.
Historically, the research of topological band insulators started
from the two dimensional (2D) quantum Hall effect.
Landau level (LL) quantization gives rise to nontrivial band
topology characterized by integer-valued Chern numbers.
\cite{thouless1982,kohmoto1985}
In fact, LLs are not the only possibility for realizing
topological band structures.
Quantum anomalous Hall band insulators with the regular Bloch-wave
structure are in the same topological class as 2D LL systems
in magnetic fields \cite{haldane1988}.
Later developments generalize the anomalous Hall insulators to time-reversal
(TR) invariant systems in both two and three dimensions.
\cite{bernevig2006,kane2005,kane2005a,bernevig2006a,fu2007,fu2007a,
moore2007,qi2008,roy2010}
This is a new class of topological band insulators with TR symmetry
which are characterized by the Z$_2$ index.
Experimentally, the most obvious signatures of band topology appear on
open boundaries, in which they exhibit helical edge or surface states.
Various 2D and 3D materials are identified as topological insulators,
and their stable helical boundary modes have been detected
\cite{konig2007,hsieh2008,zhang2009,hsieh2009,xia2009,hasan2011}.
Furthermore, systematic classifications have been performed in
topological insulators and superconductors in all the
spatial dimensions, which contain ten different universal classes
\cite{ryu2009,kitaev2009}.

Although the current research is mostly interested in topological
insulators with Bloch-wave band structures, the advantages of LLs make
them appealing for further studies.
We use the terminology of LLs here in the following general sense not just
for the usual 2D LLs in magnetic fields: {\it topological} single-particle
level structures labeled by {\it angular momentum} quantum numbers with
flat or nearly flat spectra.
On open boundaries, LL systems develop gapless surface or edge modes which
are robust against disorders.
For example, in the 2D quantum Hall LL systems, chiral edge states are
responsible for quantized charge transport.
For the 2D LL based quantum spin Hall systems, helical edge modes are
robust against TR-invariant disorders \cite{bernevig2006}.
Similar topological properties are expected for even high-dimensional
LL systems, which exhibit stable gapless surface modes.
For the usual 2D LLs, the
symmetric gauge is used in which angular momentum is conserved.
We do not use the Landau gauge because it does not maintain rotational
symmetry explicitly.
LL wavefunctions are simple and explicit, and their elegant analytical
properties nicely provide a platform for further study of
topological many-body states in high dimensions.

Generalizing LLs to high dimensions started by Zhang and Hu \cite{zhang2001}
on the compact $S^4$ sphere by coupling large spin
fermions to the SU(2) magnetic monopole, where fermion spin scales
with the radius as $R^2$.
Later on various generalizations to other manifold were developed.
\cite{karabali2002,elvang2003,bernevig2003,hasebe2010,fabinger2002}
Two of the authors have generalized the LLs of non-relativistic fermions
to arbitrary dimensional flat space $R^D$ ~\cite{li2011}.
The general strategy is very simple: the harmonic oscillator
plus spin-orbit (SO)
coupling $L_{ij} \Gamma_{ij}$, where $L_{ij}$ and $\Gamma_{ij}$ are
the orbital and spin angular momenta in a general dimension.
Reducing back to two dimensions, it becomes the quantum spin Hall Hamiltonian
in which each spin component exhibits the usual 2D LLs in the
symmetric gauge, but the chiralities are opposite for two spin
components \cite{bernevig2006a}.
For a concrete example, say, in three dimensions, each LL contributes a branch of
helical Dirac surface modes at the open boundary, thus its topology
belong to the $Z_2$-class.
Furthermore, LLs have also been constructed to arbitrary dimensional
flat spaces for relativistic fermions \cite{li2011a}, which is a square
root problem of the above non-relativistic cases.
It is a generalization of the quantum Hall effect in graphene
\cite{novoselov2005,zhang2005,castro_neto2009}
to high dimensional systems with the full rotational symmetry.
This construction can also be viewed as a generalization of the Dirac equation
from momentum space to phase space by replacing the momentum operator
with the creation and annihilation operators of phonons.
The zero-energy LL is a branch of half-fermion modes.
When it is empty or fully occupied, fermions
are pumped from the vacuum, a generalization of parity anomaly \cite{niemi1983,
jackiw1984,redlich1984,semenoff1984} to high dimensions.

In this article, we study another class of isotropic LLs with
TR symmetry but breaking parity in two and three dimensions, which can also
be straightforwardly generalized to arbitrary dimensions.
The Hamiltonians are again harmonic oscillator plus SO couplings,
but here the SO coupling is the coupling between spin and linear
momentum, not orbital momentum.
In 2D, it is simply the standard Rashba SO coupling, and in 3D
it is the $\vec \sigma \cdot \vec p$-type SO coupling.
In both cases, parity is broken.
The strong SO coupling provides the projection of the low energy
Hilbert space composed of states with the proper helicity.
The radial quantization from the harmonic potential further
generates gaps between LLs.
The SO coupling strongly suppresses the dispersion with respect
to the angular momentum within each LL.
In two and three dimensions, they exhibit gapless helical boundary modes which
are stable against TR-invariant perturbations, thus
they belong to the $Z_2$ topological class.
In fact, parent Hamiltonians, whose first LL wavefunctions are
obtained analytically and whose spectra are exactly flat,
can be constructed by the dimensional reduction method from the
high-dimensional LL Hamiltonians constructed in Ref. [\onlinecite{li2011}].

This paper is organized as follows.
The study of isotropic and TR-invariant LLs with parity breaking
is presented in Sect. \ref{sect:2D}.
The generalization to three dimensions is given in Sec. \ref{sect:3D}.
The experimental realization of the 3D Rashba-like $\vec\sigma \cdot \vec p$
-type SO coupling is performed in Sec. \ref{sect:exp}.
Conclusions and outlook are summarized in Sec. \ref{sect:conc}.


\section{Two-dimensional spin-orbit coupled Landau levels with harmonic
potential}
\label{sect:2D}

In this section, we consider the Hamiltonian of Rashba SO coupling combined
with a harmonic potential
\bea
H_{2D}=-\frac{\hbar^2\nabla^2}{2m}+\frac{1}{2}m \omega^2r^2
-\lambda (-i\hbar\nabla_x \sigma_y +i\hbar \nabla_y \sigma_x),
\label{eq:rashba}
\eea
where $\omega$ is the trapping frequency; $\lambda$ is the SO coupling
strength with the unit of velocity.
Equation (\ref{eq:rashba}) is invariant under the $SO(2)$ rotation
and the vertical-plane mirror reflection.
In other words, the system enjoys $C_{v\infty}$ symmetry.
Equation (\ref{eq:rashba}) also satisfies the TR symmetry of fermions, {\it i.e.},
$T=i\sigma_2 K$, with $T^2=-1$ and $K$ the complex conjugation.
However, parity symmetry is broken explicitly by the Rashba term.

Equation \ref{eq:rashba} can be realized in solid-state
quantum wells and ultra-cold atomic traps.
Rashba SO coupling due to inversion symmetry breaking at 2D
interfaces has been studied extensively in the
condensed matter literature; \cite{rashba1960} its energy scale
can reach very large values. \cite{ast2007}
Furthermore, Wigner crystallization in the presence of
Rashba SO coupling has been studied. \cite{berg2011}
In the context of ultra cold atoms,
Bose-Einstein condensation with Rashba SO coupling plus harmonic
potential was studied by one of the authors and Mondragon in
Ref. \onlinecite{wu2011}, in which the spontaneous generation of
a half-quantum vortex is found.
Later, there was great experimental progress in generating a
synthetic gauge field from light-atom interaction, \cite{lin2011} which
inspired a great deal of theoretical interest. \cite{ho2011,wang2010,
hu2011,ghosh2011,sinha2011,xfzhou2011}

\subsection{Energy spectra}

In a homogeneous system with Rashba SO coupling, {\it i.e.},
$\omega=0$ in Eq. (\ref{eq:rashba}), the single-particle
states $\psi_\pm (\vec k)$ are eigenstates of the helicity operator
$\vec \sigma \cdot (\vec k \times \hat z)$ with eigenvalues $\pm 1$,
respectively.
The spectra for these two branches are $\epsilon_{\pm}(\vec k)=
\hbar^2(k \mp k_{0})^2/(2m)$, and the lowest energy states are located
around a ring  with radius $k_{0}=m\lambda/\hbar$ in momentum space.
Such a system has two length scales: the characteristic length
of the harmonic trap $l_T=\sqrt{\frac{\hbar }{m\omega}}$, and the SO
length scale $l_{so}=1/k_{0}$.
The dimensionless parameter $\alpha=l_T/l_{so}$ describes the SO coupling
strength with respect to the harmonic potential.

As presented in Ref. [\onlinecite{wu2008}] for the case of strong SO
coupling, {\it i.e.}, $\alpha\gg 1$, the physics picture is
mostly clear in momentum representation.
The lowest energy states  are reorganized from the plane-wave states
$\psi_+(\vec k)$ with $\vec k$ near the SO ring.
Energetically, these states are separated from the opposite-helicity
ones $\psi_-(\vec k)$ at the order of $E_{so}=\hbar k_{0}
\lambda=\alpha^2 E_{tp}$,  where $E_{tp}=\hbar \omega$ is the
scale of the trapping energy.
As shown below, the band gap in such a system is at the scale of $E_{tp}$.
Since $\alpha\gg 1$, we can safely project out the negative helicity
states $\psi_{-}(\vec k)$.
After the projection, the harmonic potential in momentum representation
becomes Laplacian coupled to a Berry connection $\vec A_k$ as
\bea
V_{tp}=\frac{m}{2} \omega^2 (i\vec {\nabla}_k - \vec {A}_k)^2,
\eea
which drives particle moving around the ring with a moment of inertial
$I= M_k  k_{0}^2$;
$M_k=\hbar^2 /(m\omega^2)$ is the effective mass in momentum representation.
The Berry connection $A_k$ is defined as
\bea
\vec A_k=i\avg{\psi_{k+}|\vec {\nabla}_k|\psi_{k+}}=\frac{1}{2k}\hat e_{k},
\eea
where $|\psi_{k+}\rangle$ is the lower branch eigenstate with momentum
$\vec k$.
It is well known that for the Rashba Hamiltonian, the Berry connection
$A_k$ gives rise to a $\pi$ flux at $\vec k=(0,0)$ but without
Berry curvature at $\vec k\neq 0$. \cite{xiao2010}
This is because a two-component spinor after a 360$^\circ$ rotation
does not come back to itself but acquires a minus sign.

The crucial effect of the $\pi$ flux in momentum space is that
the angular momentum eigenvalues become half-integers as $j_z=m+\frac{1}{2}$.
The angular dispersion of the spectra becomes
$E_{agl}(j_z)=\hbar^2 j_z^2/2I=(j_z^2/2\alpha^2) E_{tp}$.
On the other hand, the radial potential in momentum representation is
$V(k)=\frac{1}{2}M_k \omega^2 (k-k_{0})^2$ for positive-helicity states.
For states with energies much lower than $E_{so}$, we approximate
$V(k)$ as harmonic potential, thus the radial quantization is
$E_{rad}(n_r)=(n_r+\frac{1}{2}) E_{tp}$ up to a constant.
The same dispersion structure was also noted in recent works
\cite{hu2011,sinha2011,ghosh2011}, which show
\bea
E_{n_r,j_z}\approx \Big(n_r+ \frac{1}{2}
-\frac{\alpha^2}{2}
+\frac{j_z^2}{2\alpha^2} \Big )E_{tp}
\label{eq:2D_Rashba_gap},
\eea
where the zero point energy is restored here.
Since $\alpha\gg 1$, we treat $n_r$ as a band index and $j_z$
as a good quantum number for labeling states inside each band.

\subsection{Dimensional reduction from the 3D Landau level Hamiltonian}
\label{sect:reduction_2D}

Equation (\ref{eq:rashba}) not only can be introduced from the solid-state
and cold atom physics contexts, but also can be viewed
as a result of dimensional reduction from a 3D LL Hamiltonian
[Eq. (\ref{eq:3DLL})] proposed by the authors in Ref.
\onlinecite{li2011}.
This method builds up the connection of two topological Hamiltonians in
three dimensions with inversion symmetry and two dimensons with inversion symmetry breaking.
The resultant 2D Hamiltonian Eq. (\ref{eq:2D_reduce}) exhibits
the same physics that Eq. (\ref{eq:rashba}) does for eigenstates with
$j_z<\alpha$ in the case of $\alpha\gg 1$.
The advantage of Eq. (\ref{eq:2D_reduce}) is that its lowest LL wavefunctions
are analytically solvable and their spectra are flat.

Just like the usual 2D LL Hamiltonian in the symmetric gauge, which is
equivalent to a 2D harmonic oscillator plus the orbital Zeeman term, the 3D
LL Hamiltonian is as simple as a 3D harmonic potential plus SO coupling
\cite{li2011}
\bea
H_{3D,LL}=
\frac{p^2}{2m}+\frac{1}{2}m \omega^2 r^2-\omega \vec L \cdot \vec \sigma,
\label{eq:3DLL}
\eea
which possesses 3D rotational symmetry and TR symmetry.
Its eigen-solutions are classified into positive- and
negative-helicity channels according to the eigenvalues of $\vec \sigma
\cdot \vec L=l \hbar$ or $-(l+1)\hbar$, respectively.
In the positive (negative)-helicity channel, the total angular momentum
$j_{\pm}=(l\pm\frac{1}{2})\hbar$.
The spectra in the positive-helicity channel,
$E_{n_r,l}=(2n_r+\frac{3}{2})\hbar \omega$,
are dispersionless with respect to the value of $j_+$, thus
these states are LLs.
In the presence of an open boundary, each filled LL contributes a
branch of helical Dirac Fermi surface described as
\bea
H_{sf}= v_f (\vec \sigma \times \vec p)\cdot \hat e_r -\mu,
\eea
where $\hat e_r$ is the local normal direction of the surface,
$v_f$ the Fermi velocity, and $\mu$ the chemical potential.
The stability of surface states under TR-invariant
perturbations are characterized by the $Z_2$ topological index.

Now let us perform the dimension reduction on Eq. (\ref{eq:3DLL})
by cutting a 2D off-centered plane perpendicular to the $z$-axis with
the interception $z_0$.
Within this 2D plane of $z=z_0$, Eq. (\ref{eq:3DLL}) reduces to
\bea
H_{2D,re}&=& H_{2D} - \omega L_z \sigma_z.
\label{eq:2D_reduce}
\eea
The first term is just Eq. (\ref{eq:rashba}) with Rashba SO strength
$\lambda=\omega z_0$, and the 2D harmonic trap
frequency is the same as the coefficient of the $L_z\sigma_z$ term.
The dimensionless parameter $\alpha=l_T/l_{so}=|z_0|/l_T$.
If $z_0=0$,  Rashba SO coupling vanishes.
In this case, Eq. (\ref{eq:2D_reduce}) becomes the 2D quantum spin Hall
Hamiltonian proposed in Ref. \onlinecite{bernevig2006}, which is a
double copy of the usual 2D LL with opposite chiralities
for spin-up and -down components.
At $z_0\neq 0$, Rashba coupling appears which breaks the conservation of
$\sigma_z$.

Two of the authors found the lowest LL solutions for Eq. (\ref{eq:3DLL}),
whose center is shifted from the origin to
$\vec r_c= (0,0,z_0)$ in Ref. \onlinecite{li2011}.
These states do not keep $j$ conserved but do maintain $j_z$ as a
good quantum number as
\bea
\psi_{3D,j_z,z_0}(\rho,\phi,z)&=&
e^{-\frac{\rho^2 + (z-z_0)^2}{2l_T^2}}
e^{im\phi}\nn \\
&\times&
\left( \begin{array}{c}
J_m(k_0 \rho) \\
-\sgn(z_0) e^{i\phi} J_{m+1} (k_0 \rho)
\end{array} \right),
\eea
where $\rho=\sqrt{x^2+y^2}$; $j_z=m+\frac{1}{2}$; $k_0=z_0/l_T^2$;
and $\phi$ is the azimuthal angle around the $z$ axis.
The $\psi_{3D,j_z,z_0}$'s form a complete set of the lowest LL wave functions,
but they are nonorthogonal if their $j_z$'s are the same.
By setting $z=z_0$ in the above wavefunctions, we define
the 2D reduced wave functions as
\bea
\psi_{2D,j_z}(\rho,\phi)=
e^{-\frac{\rho^2}{2l_T^2}}
\left( \begin{array}{c}
e^{im\phi} J_m(k_0 \rho) \\
-\sgn(z_0) e^{i(m+1)\phi} J_{m+1} (k_0 \rho)
\end{array} \right).~
\label{eq:2D_LL_rashba}
\nn \\
\eea
Noticing that $\partial_z \psi_{3D,j_z,z_0}|_{z=z_0}=0$,
it is straightforward to check that $\psi_{2D,j_z}$'s are solutions
for the lowest LLs for the 2D reduced Hamiltonian in Eq. (\ref{eq:2D_reduce}) as
\bea
H_{2D,re} ~\psi_{2D,j_z}= \Big(1 -\frac{\alpha^2}{2} \Big)
\hbar \omega~ \psi_{2D,j_z}.
\eea
The TR partner of Eq. (\ref{eq:2D_LL_rashba}) can be written as
\begin{widetext}
\bea
\psi_{2D,-j_z}(\rho,\phi)=
e^{-\frac{\rho^2}{2l_T^2}}
\left( \begin{array}{c}
\sgn(z_0) e^{-i(m+1)\phi}  J_{m+1}(k_0 \rho) \\
e^{-im\phi} J_{m} (k_0 \rho)
\end{array} \right)
=(-)^{m+1}\sgn(z_0) e^{-\frac{\rho^2}{2l_T^2}}
\left( \begin{array}{c}
 e^{-i(m+1)\phi} J_{-(m+1)}(k_0 \rho) \\
- \sgn(z_0)e^{-im\phi} J_{-m} (k_0 \rho)
\end{array} \right).
\eea
\end{widetext}

\subsection{Relation between Eq. (\ref{eq:rashba})
and  Eq. (\ref{eq:2D_reduce}})

\begin{figure}[tbp]
\centering\epsfig{file=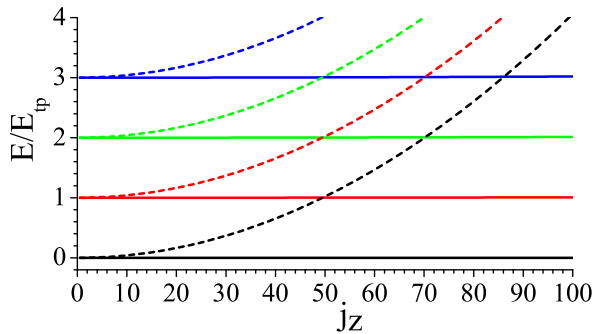,clip=1,width=0.9\linewidth, angle=0}
\caption{(Color online) Energy dispersions of the solutions for the first four LLs
to the 2D reduced Hamiltonian Eq. (\ref{eq:2D_reduce}) (solid lines), and
those for (Eq. \ref{eq:rashba}) (dashed lines).
The value of $\alpha=l_T/l_{so}=35$.
The lowest LLs of Eq. (\ref{eq:2D_reduce})
are dispersionless with respect to $j_z$.
Please note that the overall shift of the zero-point energy difference
$\frac{1}{2}\hbar \omega$ is performed for the spectra of Eq. (\ref{eq:rashba})
for a better illustration.
}
\label{fig:2DLL}
\end{figure}

The difference between the two Hamiltonians, Eq. (\ref{eq:2D_reduce}) and
(Eq. \ref{eq:rashba}), is the $L_z\sigma_z$ term.
Its effect depends on the distance $\rho$ from the center.
We are interested in the case of $|z_0|\gg l_T$, {\it i.e.}, $\alpha\gg 1$.
Let us first consider the lowest LL.
With small values of $j_z$, {\it i.e.}, $m<\alpha$,
$J_{m}(k_0\rho)$ and $J_{m+1}(k_0\rho)$ already decay before reaching
the characteristic length $l_T$ of the Gaussian factor.
We approximate their classic orbital radii as the locations of the maxima
of Bessel functions, which are roughly $\rho_{c,j_z}\approx \frac{m}{\alpha}
l_T<l_T$.
In this regime, the effect of $L_z \sigma_z$ compared to the Rashba part
is a small perturbation, of the order of $\omega \rho_{c,j_z}/
\lambda=\rho_{c,j_z}/z_0\ll 1$.
Thus, these two Hamiltonians share the same physics.
On the other hand, let us consider the case of very large values of $j_z$,
say, $m\gg \alpha^2$.
The Bessel function behaves like $\rho^m$ or $\rho^{m+1}$ at
$0<\rho<\frac{m}{\alpha} l_T$.
The classic orbit radii are just $\rho_{c,j_z}\approx \sqrt{m} l_T$.
The physics of Eq. (\ref{eq:2D_reduce}) in this regime is dominated by
the $L_z \sigma_z$ term and, thus, is the same as that of 2D quantum
spin Hall LL wave functions.
However, for Eq. (\ref{eq:rashba}), the projection to the sub-Hilbert space
spanned by $\psi_+(\vec k)$ is not valid.
Its eigenstates in this regime cannot be viewed as LLs anymore.
For intermediate values of $j_z$, i.e., $\alpha<m< \alpha^2$, the physics
is a crossover between the above two limits.

For higher LLs of Eqs. (\ref{eq:2D_reduce}) and (\ref{eq:rashba}),
we expect that their wave functions can be approximated by a
form of Eq. (\ref{eq:2D_LL_rashba}) by multiplying a polynomial of
$\rho$ at the $n_r$-th power.
As a result, the physics is similar to what is analyzed in the previous
paragraph.
At small values of $j_z<\alpha$, the energy gap is quantized in terms of
the unit of $E_{tp}=\hbar \omega$ as in Eq. (\ref{eq:2D_Rashba_gap})
for both Hamiltonians.
At very large values of $j_z\gg \alpha^2$, the LLs of
Eq. (\ref{eq:2D_reduce}) become flat again and the quantization gap is at
$2E_{tp}=2\hbar \omega$.

We perform the numerical calculation of the energy levels of the reduced 2D
Hamiltonian Eq. (\ref{eq:2D_reduce}), as plotted in Fig. \ref{fig:2DLL}.
The numerically calculated spectra of Eq. \ref{eq:rashba}, which were plotted in
Refs. \onlinecite{hu2011} and \onlinecite{sinha2011}, are also presented for comparison.
Only the spectra of $j_z>0$ are plotted, and those of $j_z<0$ are degenerate
with their partners by the TR transformation which flips the sign of $j_z$.
The lowest LL of Eq. (\ref{eq:2D_reduce}) is flat as expected, while higher LLs
are weakly dispersive which is hardly observable for the range of $j_z$
presented.
The LLs of Eq. (\ref{eq:rashba}) are dispersive with the
dependence on $j_z$ shown in Eq. (\ref{eq:2D_Rashba_gap}).
Inside the gaps between adjacent LLs of Eq. (\ref{eq:rashba}), the number
of states is of the order of $\alpha$.

\subsection{The Z$_2$ nature of the topological properties}
\label{sect:2D_topo}
Due to their connection to the 2D reduced version of the LL
Hamiltonian, we still denote the low-energy bands of Eq. (\ref{eq:rashba})
as 2D parity breaking LLs.
As shown in Eq. (\ref{eq:2D_Rashba_gap}), although these LLs are not exactly
flat, their dispersion over $j_z$ is strongly suppressed by the large
value of $\alpha$.
If the chemical potential $\mu$ lies in the middle of the band gap,
the Fermi angular momentum $j_{kf,z}$ is at the order of $\alpha$.
The classic radius of such a state is roughly $l_T$.
As analyzed in Sec. \ref{sect:reduction_2D}, for states with
$|j_z|<\alpha$, two Hamiltonians Eqs. (\ref{eq:rashba})
and (\ref{eq:2D_reduce}) share the same physics.

Compared to the usual 2D LL states, the SO coupled LLs of
Eq. (\ref{eq:rashba}) in the form of Eq. (\ref{eq:2D_LL_rashba}) are
markedly different.
The smallest length scale is not $l_T$,
but the SO coupling length scale $l_{so}=l_T/\alpha\ll l_T$.
Instead, we can use $l_T$ as the cut-off of the sample size
by imposing an open boundary condition at the radius of $l_T$.
States with $|j_z|< \alpha$ are considered as bulk states
which localize within the region of $\rho< l_T$.
States with $|j_z|\sim \alpha$ are edge states.

We take the thermodynamic limit as follows.
First, $\omega$ is fixed, which determines the LL gaps.
Then we set $m\rightarrow 0$ and $\lambda\rightarrow
\infty$ while keeping $l_{so}=\hbar/(m\lambda)$ unchanged, such that
$l_T=\sqrt{\frac{\hbar}{m\omega}}\rightarrow \infty$.
The number of bulk states scales linearly with $\alpha$,
and the level spacing scales as $1/\alpha \rightarrow 0$
at the Fermi angular momentum $j_{k_f,z}\sim \alpha$.

The next important question is the stability of the gapless edge modes.
This situation is different from the usual 2D LL problem, in which
inside each LL for each value of angular momentum $m$, there is only one state.
Those edge modes are chiral and, thus, robust against external perturbations.
Since Eq. (\ref{eq:rashba}) is TR symmetric, for each filled LL
there is always a pair of degenerate edge modes
$\psi_{n_r,\pm j_z}$ on the Fermi energy, where $n_r$ is the LL index.
Nevertheless, these two states are Kramer pairs under the TR transformation
satisfying $T^2=-1$.
In other words, the edge modes are helical rather than chiral.

We generalize the reasoning in Ref. \onlinecite{kane2005} and \onlinecite{kane2005a}
for topological insulators with good quantum numbers of
lattice momenta to our case with angular momentum good quantum numbers.
Any TR-invariant perturbation cannot mix these two states to open a gap.
In other words, the mixing term,
\bea
H_{mx}=g(\psi^\dagger_{2D,n_r,j_z} \psi_{2D,n_r, -j_z}+h.c.),
\eea
is forbidden by TR symmetry.
On the other hand, if two LLs with indices $n_r$ and $n_r^\prime$
cut the Fermi energy, the mixing term,
\bea
H_{mx}&=&g^\prime(\psi^\dagger_{2D,n_r,j_z} \psi_{2D,n^\prime_r, -j_z}
-\psi^\dagger_{2D,n^\prime_r,j_z} \psi_{2D,n_r, -j_z}\nn\\
&+&h.c.),
\eea
is allowed by TR symmetry and opens the gap.
Consequently, the topological nature of such a system is
characterized by the Z$_2$ index, even though it is not
clear how to define the Pfaffian-like formula for it due
to the lack of translational symmetry. \cite{kane2005a}
Similarly to the 2D topological insulators based on lattice
Bloch-wave states, in our case, if odd numbers of LLs
are filled such that there are odd numbers of helical edge modes,
the gapless edge modes are robust.

Imagining an open boundary at $\rho\approx l_T$, we derive an effective
edge Hamiltonian for these helical edge modes.
As $|j_z|\sim \alpha$ and taking the limit of $\alpha\rightarrow +\infty$,
these edge modes are pushed to the boundary.
We  expand the spectra around $j_{z,fm}$.
The edge Hamiltonian in the basis of $j_z$ can be written as
\bea
H_{edge}=
\sum_{j_z}  (\frac{\hbar v_f}{l_T} |j_z|-\mu) \psi^\dagger_{n_r,j_z} \psi_{n_r,j_z}
\eea
where $\mu= \frac{\hbar v_f}{l_T} j_{z,fm}$.
The edge modes $\psi_{n_r,\pm j_z}$ around $j_{z,fm}$ can also be expanded as
\bea
\psi_{n_r,j_z}&=&\left(
\begin{array}{c}
f_{n_r} e^{im\phi} \\
g_{n_r} e^{i(m+1)\phi}
\end{array}
\right), ~~
\psi_{n_r,-j_z}=T \psi_{n_r,j_z}.
\eea
$f_{n_r}$ and $g_{n_r}$ are real numbers parameterized as
\bea
f_{n_r}=\cos \frac{\theta_{n_r}}{2}, \ \ \, g_{n_r}= \sin \frac{\theta_{n_r}}{2},
\eea
which
are determined by the details of the edge.
We neglect their dependence on $|j_z|$ for states close enough to
the Fermi energy.
The effective edge Hamiltonian can also be expressed in the plane-wave
basis if we locally treat the edge as flat
\bea
H_{n_r,edge}&=& v_f\Big (\sin\theta_{n_r}
[(\vec p \times \hat e_r) \cdot \hat z )]
(\vec \sigma \cdot \hat e_r) \nn \\
&+& \cos\theta_{n_r}
(\vec p \times \hat e_r)\cdot \sigma_z \Big)  -\mu,
\eea
where $\hat e_r$ is the local normal direction on the circular edge;
both terms are allowed by rotational symmetry, TR symmetry, and
the vertical mirror symmetry in such a system.
Each edge channel is a branch of helical one-dimensional Dirac fermion modes.

Equation (\ref{eq:rashba}) can be defined on the compact $S^2$ sphere, which
takes the simple form
\bea
H=\frac{L^2}{2 I}-\omega \vec L \cdot \vec \sigma.
\eea
The eigenvalues of $\vec L \cdot \vec \sigma$ take $l \hbar$
and $-(l+1) \hbar$ for the positive and negative helicities of
$j_\pm =l\pm \frac{1}{2}$, respectively.
For convenience, we choose the parameter value of $I\omega/\hbar$
as a large half-integer, then for the lower energy branch, the energy
minimum takes place at $j_{0,+}=l_0+\frac{1}{2}=I\omega/\hbar$.
The lowest LLs become SO-coupled harmonics with $j_+=j_{0,+}$
and ($2l_0+2$)-fold degeneracy.
The gap between the lowest LLs and higher LLs is $\Delta =\hbar^2/(2I)$,
which is independent of $\omega$.
To take the thermodynamic limit, we keep $I$ constant while increasing
the sphere radius $R$, and maintain $\omega$ scaling with $R^2$,
such that the density of states on the sphere is a constant.


\section{Three-dimensional spin-orbit $\vec \sigma\cdot \vec p$ coupling in the harmonic trap}
\label{sect:3D}

In this section, we generalize the results in Sec. \ref{sect:2D}
to three dimensions.
We consider the $\vec\sigma \cdot \vec p$-type SO coupling combined
with a 3D harmonic potential
\bea
H_{3D}=-\frac{\hbar^2 \nabla^2}{2m}+\frac{1}{2}m \omega^2r^2
-\lambda (-i\hbar\vec \nabla \cdot \vec \sigma).
\label{eq:3DSO}
\eea
Equation (\ref{eq:3DSO}) possesses the 3D rotational symmetry, and
TR symmetry of fermions with $T^2=-1$.
The parity symmetry is broken by the $\vec \sigma \cdot \vec k$ term,
and there is no mirror plane symmetry either.
The quantities $l_{so}$, $l_T$, $\alpha$, and $k_{0}$ are defined
in the same way as in Sec. \ref{sect:2D}.

Although it is difficult to realize strong SO coupling
in the form of $\sigma\cdot \vec p$ in solid-state systems, it can
be designed through light-atom interactions in ultra cold atom
systems.
We present an experimental scheme to realize Eq. (\ref{eq:3DSO})
in Sec. \ref{sect:exp}.

\subsection{Energy spectra}

Again, we consider the limit of strong SO coupling, {\it i.e.}, $\alpha\gg 1$.
It is straightforward to generalize the momentum space picture in Sec.
\ref{sect:2D} to the 3D case as presented in Ref. \onlinecite{ghosh2011}
and summarized below.
The helicity operator $\vec \sigma \cdot \hat k$ is employed to define
the helicity eigenstates of plane waves $(\vec \sigma
\cdot \hat k) \psi_{\vec k,\pm}=\pm \psi_{\vec k,\pm}$.
Only positive-helicity states $\psi_{\vec k+}$ are kept in the low
energy Hilbert space.
The harmonic potential becomes the Laplacian operator in momentum space, and
thus is equivalent to a quantum rotor subject to the Berry phase in
momentum space as $V_{tp}=\frac{1}{2}m (i \vec{\nabla}_k -\vec A_k)^2$.
The moment of inertial is again $I=M_k k_{0}^2$ and $M_k=\hbar^2/(m\omega^2)$.
The Berry connection
$\vec A_k=i\avg{\psi_{\vec k, \pm}|\vec{\nabla}_k|\psi_{\vec k, \pm}}$ is the vector
potential of the U(1) magnetic monopole.
As a result, the angular momentum quantization changes to that $j$
takes half-integer values starting from $\frac{1}{2}$.
The energy dispersion becomes $E_{agl}(j)=\hbar^2 j(j+1)/2I=
(j(j+1)/2\alpha^2) E_{tp}$, and each level is ($2j+1$)-fold degenerate.
The radial quantization is the same as before.
Thus the dispersion can be summarized as
\bea
E_{n_r,j,j_z}\approx \Big(n_r+ \frac{1}{2}-\frac{\alpha^2}{2}
+\frac{j(j+1)}{2\alpha^2} \Big )E_{tp}
\label{eq:3D_SO_gap},
\eea
where $n_r$ is the band index, or, the LL index,  and $j$
is the angular momentum quantum number.

\subsection{Dimensional reduction from the 4D Landau level Hamiltonian}
\label{sect:reduction_3D}

Following the same logic as in Sec. \ref{sect:reduction_2D}, we present the
dimensional reduction from the 4D LL Hamiltonian [Eq. (\ref{eq:4DLL})] to arrive at a 3D SO
coupled Hamiltonian closely related to Eq. (\ref{eq:3DSO}).
The 3D LL Hamiltonian, Eq. (\ref{eq:3DLL}), can be easily generalized to
arbitrary dimensions by combining the $n$-D harmonic potential and the
$n$-D SO coupling between orbital angular momenta and
fermion spins in the fundamental spinor representations. \cite{li2011}
In four dimensons, there are two non-equivalent fundamental spinors, both of
which have two components.
Without loss of generality, we choose one of them as
\bea
\sigma_{ij}=\epsilon_{ijk}\sigma_k,  \ \ \
\sigma_{i4}=\sigma_i,
\eea
where $i,j=1,2$, and $3$.
The orbital angular momentum operators are defined as
$L_{ij}=-i\hbar x_i \nabla_j +i\hbar x_j \nabla_i$ where
$i,j=1,2,3$, and $4$.
The 4D LL Hamiltonian in the flat space is defined as
\bea
H_{4D,LL}&=&\sum_{i=1}^4 -\frac{\hbar^2\nabla_i^2}{2m}+\frac{m\omega^2}{2}
\sum_{i=1}^4 r_i^2 \nn \\
&-&\omega \sum_{1\le i<j\le 4} L_{ij} \sigma_{ij},
\label{eq:4DLL}
\eea
which possesses TR and parity symmetry.

The $l$th-order 4D orbital spherical harmonics coupled to the
fundamental spinor can be decomposed into the 4D SO-coupled spherical
harmonics in the positive- and negative-helicity sectors, where
$L_{ij}\sigma_{ij}$ take eigenvalues of $l \hbar$ and $-(l+2) \hbar$, respectively.
The eigen wave functions of Eq. (\ref{eq:4DLL}) in the positive-helicity
channel are dispersionless with respect to $l$ as $E_{n_r, +}
=(2n_{r}+2)\hbar \omega$.
Their radial wave functions are $R_{n_{r},l}(r)= r^{l}e^{-r^{2}/2l_{T}^{2}}
F(-n_{r},l+2,r^{2}/l_{T}^{2})$, where $F$ is the standard
confluent hypergeometric function.
With an open boundary of an $S^3$ sphere, each filled LL
contributes to a gapless surface mode of 3D Weyl fermions as
\bea
H_{3D, surface}= v_f \hat e_{r,i} \sigma_{ij} p_j -\mu,
\eea
where $\hat e_r$ is the unit vector normal to the $S^3$
sphere.
The topological index for such a 4D LL systems with TR symmetry
is $Z$ rather than $Z_2$.

We perform the dimensional reduction on Eq. (\ref{eq:4DLL}) from four to  three dimensions.
We cut a 3D off-center hyper-plane perpendicular to the fourth axis with the
interception $x_4=w_0$
Within this 3D hyper-plane of $(x_1,x_2,x_3, x_4=w_0)$,
Eq. (\ref{eq:4DLL}) reduces to
\bea
H_{3D,redc}=H_{3D,SO}-\omega \vec L \cdot \vec \sigma,
\label{eq:3D_reduce}
\eea
where the first term is just Eq. (\ref{eq:4DLL}) with the
SO-coupling strength $\lambda=\omega w_0$.
It contains another SO-coupling term, $\vec L \cdot \vec \sigma$,
and its coefficient is the same as the harmonic trapping frequency.
Similarly to the previous reduction from three to two dimensions, here we have
$\alpha=l_T/l_{so}=|w_0|/l_T$.
At $w_0=0$, Eq. (\ref{eq:3D_reduce}) becomes the 3D LL Hamiltonian
of Eq. (\ref{eq:3DLL}) with parity symmetry.
If $w_0\neq 0$, the $\vec \sigma \cdot \vec p$ term breaks parity symmetry.
Following the same reasoning as in Sec. \ref{sect:reduction_2D},
Eqs. (\ref{eq:3DSO}) and (\ref{eq:3D_reduce}) share the same physics
for eigenstates with $j< \alpha$ in the case of $\alpha\gg 1$.

Similarly as before, we construct an off-center
solution to the 4D LL problem.
We use $\vec r$ to denote a point in the subspace of $x_{1,2,3}$,
and $\hat \Omega$ as an arbitrary unit vector in the $x_1$-$x_2$-$x_3$ space.
We consider the plane of $\hat \Omega$-$\hat x_4$ spanned by the orthogonal
vectors  $\hat\Omega$ and $\hat x_4$.
It is easy to check that the following wave functions, which depends only
on coordinates in the $\hat\Omega$-$\hat x_4$ plane are the lowest LL
solutions to the 4D LL Hamiltonian, Eq. (\ref{eq:4DLL})
\bea
(\vec r \cdot \hat \Omega + i x_4)^l e^{-\frac{r^2+x_4^2}{2l_T^2}}
\otimes \alpha_{\hat\Omega},
\label{eq:HWS}
\eea
where  $\alpha_{\hat\Omega}=(\cos\frac{\theta}{2},\sin\frac{\theta}{2}
e^{i\phi})^T$ satisfies
\bea
(\sigma_{i4}\Omega_i)
\alpha_{\hat\Omega } =(\vec \sigma \cdot \hat \Omega) \alpha_{\hat\Omega}
=\alpha_{\hat\Omega}.
\eea
In this set of wavefunctions, both the orbital angular momentum
and spin  are conserved and added up; they are called the highest weight states in group theory.
In fact, these states can be rotated into any plane accompanied by a
simultaneous rotation in the spin channel.
Based on the structure of the highest weight states, we can still
define the magnetic translation operator in the $\hat\Omega$-$x_4$ plane
along the $x_4$ axis as
\bea
T_{\hat\Omega x_4} (w_0 \hat x_4)=\exp\Big(-w_0 \partial_{x_4} -\frac{i}{l_T^2}
(\vec r \cdot \Omega) w_0 \Big).
\eea
Applying this operator to the Gaussian pocket of the solution with
$l=0$ in Eq. (\ref{eq:HWS}), we arrive at the off-center solution
\bea
\psi_{\Omega,w_0}(\vec r, x_4)= e^{-\frac{r^2+x_4^2}{2l_T^2}}
e^{-i\frac{r w_0}{l_T^2}} \otimes \alpha_{\hat\Omega}.
\eea
This solution, however, breaks the rotational
symmetry.
In order to restore the 3D rotational symmetry around the new center
$(0,0,0, w_0)$, we perform a Fourier transformation over the direction
of $\Omega$ as
\bea
\psi_{4D;j,j_z}(\vec r, x_4)&=& \int d \Omega ~
{\cal Y}_{-\frac{1}{2},l+\frac{1}{2}, m+\frac{1}{2}} (\hat \Omega)
\psi_{\Omega,w_0}(\vec r, x_4).  \nn \\
\label{eq:4D_offcenter}
\eea
where $j=l+\frac{1}{2}$ and $j_z=m+\frac{1}{2}$.
Please note that due to the singularity of $\alpha_\Omega$ over the
direction of $\hat \Omega$, monopole spherical harmonics,
${\cal Y}_{-\frac{1}{2},l+\frac{1}{2},m+\frac{1}{2}}(\Omega)$, are used instead of
regular spherical harmonics.

Again, noting that $\partial_{x_4}\psi_{4D;j,j_z}(\vec r, x_4)|_{x_4,w_0}=0$,
we simply set $x_4=w_0$; then it is simple to check that the reduced 3D
wave functions
\bea
\psi_{3D,j,j_z}(\vec r)=\psi_{4D;j,j_z}(\vec r, w_0)
\eea
are the solutions to Eq. (\ref{eq:3D_reduce}) for the lowest LLs
as
\bea
H_{3D,redc} \psi_{3D,j,j_z}(\vec r)= \Big (\frac{3}{2}
-\frac{\alpha^2}{2} \Big )\hbar \omega
\psi_{3D,j,j_z}(\vec r). \nn \\
\eea
$\psi_{3D,j,j_z}(\vec r)$ can be simplified as
\bea
\psi_{3D, jj_z}(\vec r)&=& e^{-\frac{r^2}{2l_T^2}} \Big\{ j_l(k_0 r)
Y_{+,j,l,j_z} (\Omega_r)
+i j_{l+1}(k_0 r) \nn \\
&\times& Y_{-,j, l+1, j_z} (\Omega_r) \Big\},
\label{eq:3D_WF}
\eea
where $k_0=w_0/l_T^2=m \lambda/\hbar$ and $\lambda= w_0 \omega$;
$j_l$ is the $l$th-order spherical Bessel function.
$Y_{\pm,j, l,j_z}$'s are the SO-coupled spherical harmonics defined as
\bea
Y_{+,j,l,j_z}(\Omega)=\Big(\sqrt{\frac{l+m+1}{2l+1}} Y_{lm},
\sqrt{\frac{l-m}{2l+1}} Y_{l,m+1}\Big)^T
\nn
\eea
with a positive
eigenvalue of $l\hbar$ for $\vec \sigma \cdot \vec L$, and
\bea
Y_{-,j,l,j_z}(\Omega)=\Big(-\sqrt{\frac{l-m}{2l+1}} Y_{lm},
\sqrt{\frac{l+m+1}{2l+1}} Y_{l,m+1}\Big)^T
\nn
\eea
with a negative
eigenvalue of $-(l+1)\hbar$  for $\vec \sigma \cdot \vec L$.

The difference between Eq. (\ref{eq:3D_reduce}) and Eq. (\ref{eq:3DSO})
is the term $\vec \sigma \cdot \vec L$, whose effect is weakened
as the distance from center $r$ gets small.
The radial distributions of $j_l(k_0 r)$ in Eq. (\ref{eq:3D_WF}) and
$J_m(k_0\rho)$ in Eq. (\ref{eq:2D_LL_rashba}) are similar.
Following the same reasoning presented in Sec. \ref{sect:reduction_2D},
in the limit of $\alpha\gg 1$, we can divide the lowest LL states
of Eq. (\ref{eq:3D_WF}) into three regimes: $j<\alpha$, $j\gg\alpha^2$,
and $\alpha<j<\alpha^2$.
At $j<\alpha$, the classic orbit radius scales as $r_{c,j}\approx
\frac{j}{\alpha}l_T<l_T$.
Again in this regime, the effect of $\vec \sigma \cdot \vec L$ is
a perturbation of the order of $r_{c,j_z}/z_0\ll 1$; thus the two Hamiltonians, Eq. (\ref{eq:3D_reduce}) and Eq. (\ref{eq:3DSO}), share the same physics.
Similarly, in the regime of $j\gg \alpha^2$, $\vec \sigma \cdot \vec L$
dominates, and the physics of Eq. (\ref{eq:3D_reduce}) comes back to the
3D LL Hamiltonian, Eq. (\ref{eq:3DLL}), while that of Eq. (\ref{eq:3DSO})
is no longer LL-like.

\subsection{The Z$_2$ helical surface states}
\label{sect:3D_topo}

Following the same reasoning as in Sec. \ref{sect:2D_topo}, we denote the
low-energy bands of Eq. (\ref{eq:3DSO}) as 3D parity breaking LLs.
For the lowest LL, below the energy of the bottom of the second LL,
the angular momentum $j$ takes values from $\frac{1}{2}$ to the order
of $\alpha$ at which the radius of the LL approaches $l_T$.
For this regime $j<\alpha$, Eqs. (\ref{eq:3DSO}) and (\ref{eq:3D_reduce})
share the same physics.
Again, the smallest length scale is the SO coupling length scale
$l_{so}=l_T/\alpha\ll l_T$.
States with $|j|\ll \alpha$ are considered bulk states
which localize within the region $\rho\ll l_T$.
States with $|j_z|\sim \alpha$ are edge states.
The number of bulk states scales linearly with $\alpha^2$.

Now we impose an open boundary condition of an $S^2$ sphere with radius
$r\approx l_T$, and consider the stability of the edge modes against TR
invariant perturbations.
Let us consider one filled LL.
The Fermi energy lies between the gap, and thus cuts the dispersion
at surface states.
In the limit of $\alpha\rightarrow\infty$, the energy level spacing
between adjacent angular momenta $j$ and $j+1$ scales
as $\hbar \omega/\alpha\rightarrow 0$ for surface modes with
$j\sim \alpha$.
Thus we can always choose the Fermi angular momentum $j_{f}$ satisfying
$j_{f}=2l+\frac{1}{2}$.
For this value of $j_{f}$, there is an odd number of $2l+1$ Kramer
pairs between $\psi_{j_{f}, \pm j_z}$ for $j_z=\frac{1}{2}$ to $j_{f}$.
Again according to the reasoning of the $Z_2$-classification in Refs.
\onlinecite{kane2005} and \onlinecite{kane2005a}, these states cannot be fully
gapped out by applying TR invariant perturbations.
Certainly, for those states with $j=2l+\frac{3}{2}$ close to the
Fermi energy, they can be fully gapped, but they are only part
of the spectra, and do not change the topological properties.
Again, if two LLs with different indices $n_r$ and $n_r^\prime$
cut the Fermi energy, the zero energy states at the Fermi level
can be fully gapped out.
Thus, the topological nature of Eq. (\ref{eq:3DSO}) is $Z_2$.

We further present the effective surface Hamiltonian for surface
modes in the limit of $j_{f}\sim \alpha \rightarrow +\infty$.
The effective surface Hamiltonian of the 3D topological insulators
with the spherical boundary condition has also been discussed
in Refs. \onlinecite{lee2009} and \onlinecite{parente2011}.
The surface Hamiltonian in the eigen-basis of $j$ and $j_z$ can be written as
\bea
H_{sf}=
\sum_{j,j_z}  (\frac{\hbar v_f}{l_T} |j|-\mu) \psi^\dagger_{n_r,j,j_z}
\psi_{n_r,j,j_z},
\eea
where $\mu= \frac{\hbar v_f}{l_T} j_{f}$.
The construction of the accurate surface Hamiltonian in the plane-wave
basis depends on the detailed information of surface modes
$\psi_{j,j_z}(r,\Omega_r)$ for $j\approx j_{f}$ and, thus, is cumbersome.
Nevertheless based on the symmetry analysis, we can write
the general form as
\bea
H_{n_r,edge}&=&v_f \Big\{ \sin\theta_{n_r} (\vec p \times \vec \sigma)
\cdot \hat e_r \nn \\
&+&\cos\theta_{n_r} \big[\vec p \cdot \vec \sigma
-(\vec p \cdot \hat e_r) (\vec \sigma \cdot \hat e_r)\big]
\Big\}-\mu.
\label{eq:3D_surface}
\eea
where $\hat e_r$ is the local norm direction on the $S^2$-sphere.
Both terms obey the local $SO(2)$ rotational symmetry around the
$\hat e_r$ and TR symmetry.
The first Rashba term also obeys the vertical mirror symmetry, while
the second term does not.
The second term favors the spin aligning with the momentum, while the
second favors a relative angle of 90$^\circ$.
For a general value of $\theta_{n_r}$, which is determined by the
non-universal surface properties and $\theta_{n_r}$,
Eq. (\ref{eq:3D_surface}) determines a relative rotation
between spin and momentum orientation at the angle of $\theta_{n_r}$.
It is still a helical Dirac Fermi surface.


\section{Experimental realization for 3D SO coupling}
\label{sect:exp}

\begin{figure}
\centering\epsfig{file=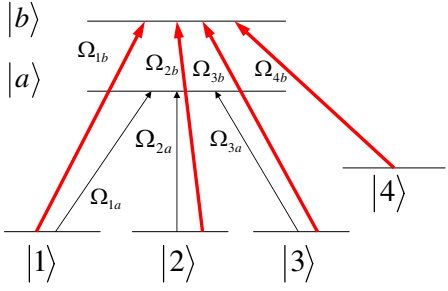,clip=1,width=0.7\linewidth, angle=0}
\caption{(Color online) Level diagram for atom-laser coupling.
Four lower energy levels are coupled to two excited levels to compose a
hybrid tripod and tetrapod configuration.
}
\label{fig:level-strct}
\end{figure}

In the ultra cold atom context, there has been great progress in
the synthetic gauge field, or, artificial SO coupling
from light-atom interactions \cite{dalibard2011}.
Experimentally, artificial SO coupling has been generate in ultra cold
atom systems. \cite{lin2011}
Two dimensional Rashba and Dresselhaus SO coupling in the harmonic
potential has been proposed using a double-tripod configuration.
\cite{juzeliunas2011}
Since the pseudo-spin degrees of freedom are represented by the two lowest energy
levels, this scheme is immune to decay due to collision and spontaneous
emission process. \cite{ruseckas2005}

In this section, we propose the experimental realization for the 3D SO coupling
of the $\vec\sigma \cdot \vec p$ type in Eq. (\ref{eq:3DSO}).
Here we generalize the scheme in Ref. \onlinecite{juzeliunas2011}
to a combined tripod and tetrapod level configuration as depicted
in Fig. \ref{fig:level-strct}.
Three internal levels $|1\rangle$, $|2\rangle$, and $|3\rangle$ couple the
excited state $|a\rangle$ to form a tripod configuration.
A tetrapod-like coupling is formed by coupling the four levels
$|1\rangle-|4\rangle$ to the common excited state $|b\rangle$.
The single-particle Hamiltonian reads
\bea
H=\frac{p ^2}{2m} + \frac{1}{2} m \omega^2  r^2 + H_{al},
\eea
where $m$ is the mass of the atom; $H_{al}$ represents the atom-laser
coupling.
In the interaction picture, $H_{al}$ can be written under the rotating
wave approximation as
\bea
H_{al}&=& - \hbar \sum_{m=a,b} \big\{ \Omega_{1m} |m\rangle \langle 1|
+ \Omega_{2m} |m \rangle \langle 2 | + \Omega_{3m} | m\rangle \langle 3|
\nonumber \\
&+& h.c.\big\}
-  \hbar \left [ \Omega_{4b} |b\rangle \langle 4| + h.c. \right ] ,
\eea
where $\Omega_{im}$ are the corresponding Rabi frequencies between the
internal states $|i\rangle$ and $|m\rangle$ with $m=a,b$ .

\begin{figure}
\centering\epsfig{file=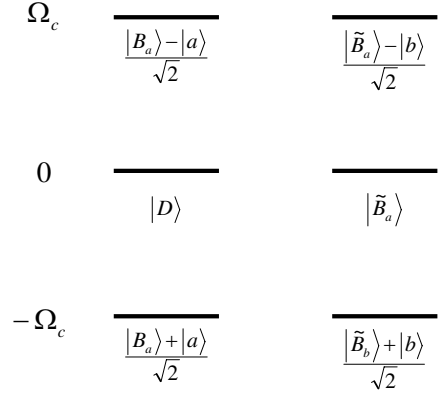,clip=1,width=0.7\linewidth, angle=0}
\caption{Energy levels of the atom-laser coupling Hamiltonian
Eq. \ref{reducedH}.
}
\label{fig:level-strct2}
\end{figure}

We introduce the following two bright states
\bea
|B_m \rangle = ( \Omega_{1m}^* |1 \rangle + \Omega_{2m}^* |2 \rangle
+ \Omega_{3m}^* |3 \rangle )/\Omega_m,
\eea
where $m=a,b$ and $\Omega_m =
\sqrt{|\Omega_{1m}|^2+|\Omega_{2m}|^2+|\Omega_{3m}|^2}$.
The atom-laser coupling can be rewritten as
\bea
H_{al} &=& - \hbar \Big\{ \Omega_{a} |a\rangle \langle B_a|
+ \emph{h.c.} \Big\} \nonumber \\
&-& \hbar \Big\{ \Omega_{b} |b\rangle \langle B_b| +
\Omega_{4b} |b\rangle \langle 4| + h.c. \Big\}.
\eea
To further simplify the model, here we assume $\langle B_a| B_b \rangle=0$,
which can be achieved by choosing
\bea
\Omega_{jm} &=& \frac{\Omega_m}{\sqrt{3}} e^{i (\vec{k}_j \cdot \vec{r} +
\theta_{jm}) }, \ \ \, (j=1,2,3; m=a,b)
\eea
with $\theta_{ja}=(j-2)2\pi/3$ and $\theta_{jb}=-(j-2)2\pi/3$.
We also choose $\Omega_{4b} = \Omega_4 e^{i (\vec{k}_4 \cdot \vec{r} + \theta_4) }$,
and set $\Omega_c=\Omega_a$, $\Omega_b = \Omega_c \cos \phi$, and
$\Omega_4 = \Omega_c \sin \phi$.
Using these notations, $H_{al}$ is simplified as
\bea
\label{reducedH}
H_{al} = - \hbar \left[\Omega_{c} (|a\rangle \langle B_a|
+ |b\rangle \langle \tilde{B}_b| )+ \emph{h.c.} \right],
\eea
where
$|\tilde{B}_b\rangle = \cos \phi | B_b \rangle + \sin \phi
|\tilde{4}\rangle$ and $|\tilde{4}\rangle = e^{-i (\vec{k}_4 \cdot \vec{r}
 + \theta_4)} | 4 \rangle$.
The above Hamiltonian supports three pairs of degenerated eigenstates
with energy difference $\hbar |\Omega_c|$, as depicted in
Fig. (\ref{fig:level-strct2}).
Explicitly, the eigen-vectors are written as
\bea
|G_1\rangle &=& \frac{| B_a\rangle + | a\rangle}{\sqrt{2}}, \ \ \,
|G_2\rangle = \frac{| \tilde{B}_b\rangle + | b\rangle}{\sqrt{2}}, \nn \\
|G_3\rangle &=&  | D\rangle, \ \ \, \ \ \, \ \ \, \ \ \,
|G_4\rangle = | \tilde{B}^{\perp}_b\rangle , \nn \\
|G_5\rangle &=&  \frac{| B_a\rangle - | a\rangle}{\sqrt{2}} , \ \ \,
|G_6\rangle = \frac{ |\tilde{B}_b\rangle + | b\rangle}{\sqrt{2}},
\eea
where $|D\rangle = \sum_j e^{-i \vec{k}_j \cdot \vec{r}} |j\rangle/\sqrt{3}$ and
$| \tilde{B}^{\perp}_b\rangle = \sin \phi | B_b \rangle - \cos \phi
|\tilde{4}\rangle $. Therefore, the two degenerate ground states can
be used as pseudo-spin $1/2$ degrees of freedom.

If the trapping frequency satisfies $\omega \ll |\Omega_c|$, according
to the adiabatic approximation, we neglect the coupling between the
ground-state manifold and other states.
Therefore, atoms in the subspace spanned by $|G_1\rangle$ and
$|G_2\rangle$ evolve according to the effective Hamiltonian
\bea
H_e=\frac{(\vec p-\vec A)^2}{2m} + \frac{1}{2}m\omega^2 r^2+\Phi,
\eea
where the non-Abelian gauge potential $\vec A$ is a $2 \times 2$ matrix
with the elements
\bea
\vec A_{ij}= i \hbar \langle G_i| \vec \nabla | G_j \rangle,
\eea
where $i,j=1,2$; $\Phi$ is a scalar potential induced by the coupling
laser beams.

An isotropic $3$D $\vec \sigma \cdot \vec p$-like SO coupling can be
obtained by a 3D set-up of laser configurations as
\bea
\vec{k}_1&=&\kappa(-\frac{1}{2},-\frac{\sqrt{3}}{2},0), \ \ \,
\vec{k}_2=\kappa(0,1,0), \nn \\
\vec{k}_3&=&\kappa(-\frac{1}{2}, \frac{\sqrt{3}}{2},0), \ \ \,
\vec{k}_4=\kappa(0,0,-\sqrt{\frac{7+\sqrt{17}}{8}}).
\eea
In this case, the corresponding vector and scale potential are
calculated as
\bea
\frac{\vec A}{\hbar} &=&  0.166 \kappa
\left [ \sigma_x \vec{e}_x + \sigma_y
\vec{e}_y +(\sigma_z -I ) \vec{e}_z \right ], \nn \\
\Phi &=& 0.445 \frac{ \hbar^2 \kappa^2}{2 m } \hat{I}.
\label{eq:wavevectors}
\eea
The $\Phi$ term is a constant and, thus, can be dropped off.
The Abelian part in the gauge potential $A_z$ is a constant,
which can be absorbed by a gauge transformation.
Consequently, the remaining constant non-Abelian gauge potential
behaves as a $\vec \sigma \cdot \vec p$ type SO coupling.

\begin{figure}
\centering\epsfig{file=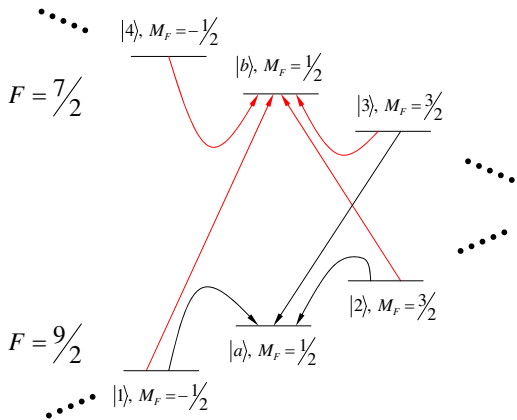,clip=1,width=0.8\linewidth, angle=0}
\caption{(Color online) Energy level scheme for alkali atoms $^{40}K$.
The Zeeman sublevels of two hyperfine states $F=\frac{9}{2}$ and
$F=\frac{7}{2}$ can be used to fulfill our requirements.
Lines or curves with an arrow indicate effective transitions
between different magnetic levels which can be implemented using resonant
Raman processes.
Other levels, which are not involved in the scheme, are not shown.}
\label{fig:level-atom}
\end{figure}

The above-considered level structure can be found for example,
in alkali atoms with large spins.
Figure \ref{fig:level-atom} shows the hyperfine ground state manifolds of
$^2S_{1/2}$ for $^{40}K$ atoms under an external magnetic field.
The energy levels $|1\rangle \sim |4\rangle$, $|a\rangle$, and $|b\rangle$
can be selected as different Zeeman sublevels of
$F=\frac{9}{2}$ and $F=\frac{7}{2}$.
 Using the notation of $|FM_F\rangle$ to denote each state, we choose
$|1\rangle = |\frac{9}{2}, -\frac{1}{2} \rangle$,
$|2\rangle = |\frac{9}{2},  \frac{3}{2} \rangle$,
$|3\rangle = |\frac{7}{2},  \frac{3}{2}\rangle$,
$|4\rangle = |\frac{7}{2}, -\frac{1}{2} \rangle$,
$|a\rangle = |\frac{9}{2},  \frac{1}{2} \rangle$ and
$|b\rangle = |\frac{7}{2},  \frac{1}{2} \rangle$.
The coupling between different levels is achieved for example, by using
two laser beams under second-order
resonant Raman process. The two lasers can be chosen to be circularly polarized
and $\pi$ polarized, respectively, in order to satisfy the selection
rule. Finally, wave vectors of individual laser beams can also be
adjusted so that Eq. (\ref{eq:wavevectors}) is fulfilled.

\section{Conclusion and outlook}
\label{sect:conc}

We have studied rotationally and TR symmetric LL systems in both 2D and 3D
systems with breaking parity symmetry, whose topological properties are
characterized by the $Z_2$ class.
These Hamiltonians are simply 2D harmonic potentials plus Rashba SO
coupling, or 3D harmonic potentials plus $\vec \sigma\cdot \vec p$-type SO coupling with a strong SO coupling strength.
For low-energy bands, the dispersions over angular momenta are
strongly suppressed by SO coupling, to be nearly flat.
Up to a small difference which can be treated perturbatively,
these Hamiltonians can be systematically investigated through dimensional
reduction on the high-dimensional LL problems
by cutting an off-center plane in the 3D LL
Hamiltonian or an off-center hyper plane in the 4D LL Hamiltonian.
The parity breaking LL wavefunctions in two and three dimensions are presented
explicitly.
With open boundary conditions, helical edge states are found in two dimensions, and
surface states are found in three dimensions.
These states can be realized in ultra cold atom systems in a
harmonic trap combined with synthetic gauge fields, {\it i.e.},
artificial SO coupling.
In particular, we propose an experimental scheme to realize the 3D
Hamiltonian.

The above dimensional procedure can be straightforwardly generalized
to arbitrary dimensions based on our previous construction of high
dimensional LL Hamiltonians, \cite{li2011} and so can
the general parity breaking LL wavefunctions in $N$ dimensions.
The nice analytical properties of the 2D and 3D LL wave functions
breaking parity symmetry also provide a good opportunity to
further construct many-body wave functions of the factional topological
states.
These properties will be investigated in a future publication.

\acknowledgments
C. W. thanks L. Balents and S. Ryu for early collaboration
on a related work. 
Y. L. and C. W. were supported by the AFOSR YIP program and
NSF Grant No. DMR-1105945.
X. F. Z. acknowledges support from the NFRP (Grant No. 2011CB921204), NNSF (Grant No. 60921091),
NSFC (Grant No. 11004186), CUSF, and SRFDP (Grant No. 20103402120031).


\begin{thebibliography}{10}

\bibitem{hasan2010}
M. Hasan and C. Kane, Rev. Mod. Phys. {\bf 82},  3045  (2010).


\bibitem{qi2011}
X.-L. {Qi} and S.-C. {Zhang}, Rev. Mod. Phys. {\bf 83}, 1057–1110 (2011).


\bibitem{thouless1982}
D.~J. Thouless, M. Kohmoto, M.~P. Nightingale, and M. den~Nijs, Phys. Rev. Lett. {\bf
  49},  405  (1982).

\bibitem{kohmoto1985}
M. Kohmoto, Ann. Phys. {\bf 160},  343  (1985).

\bibitem{haldane1988}
F.~D.~M. Haldane, Phys. Rev. Lett. {\bf 61},  2015  (1988).

\bibitem{kane2005}
C.~L. Kane and E.~J. Mele, Phys. Rev. Lett. {\bf 95},  146802  (2005).

\bibitem{kane2005a}
C.~L. Kane and E.~J. Mele, Phys. Rev. Lett. {\bf 95},  226801  (2005).

\bibitem{bernevig2006a}
B. A. Bernevig and S. C. Zhang, Phys. Rev. Lett. 96, 106802 (2006).

\bibitem{bernevig2006}
B. Bernevig, T. Hughes, and S. Zhang, Science {\bf 314},  1757  (2006).

\bibitem{fu2007}
L. Fu and C.~L. Kane, Phys. Rev. B {\bf 76},  045302  (2007).

\bibitem{fu2007a}
L. Fu, C.~L. Kane, and E.~J. Mele, Phys. Rev. Lett. {\bf 98},  106803  (2007).

\bibitem{moore2007}
J.~E. Moore and L. Balents, Phys. Rev. B {\bf 75},  121306  (2007).

\bibitem{qi2008}
X.-L. Qi, T.~L. Hughes, and S.-C. Zhang, Phys. Rev. B {\bf 78},  195424  (2008).

\bibitem{roy2010}
R. Roy, New J. Phys. {\bf 12},  065009  (2010).

\bibitem{hsieh2008}
D. Hsieh {\it et~al.}, Nature {\bf 452},  970  (2008).


\bibitem{konig2007}
M. K\"onig {\it et~al.}, Science {\bf 318},  766  (2007).


\bibitem{zhang2009}
H. Zhang {\it et~al.}, Nature Phys. {\bf 5},  438  (2009).

\bibitem{hsieh2009}
D. Hsieh {\it et~al.}, Phys. Rev. Lett. {\bf 103},  146401  (2009).

\bibitem{xia2009}
Y. Xia {\it et~al.}, Nature Phys. {\bf 5},  398  (2009).

\bibitem{hasan2011}
M. Zahid Hasan, J. E. Moore, Ann. Rev. Cond. Matt. Phys., {\bf 2}, 55 (2011).

\bibitem{ryu2009}
S. Ryu, A. Schnyder, A. Furusaki, A. Ludwig, New J. Phys. 12, 065010 (2010).

\bibitem{kitaev2009}
A. Kitaev, {\em American Institute of Physics Conference Series}, Vol. 1134, 22 (2009).


\bibitem{zhang2001}
S. C. Zhang and J. Hu, Science {\bf 294},  823  (2001).


\bibitem{karabali2002}
D. Karabali and V. P. Nair, Nucl. Phys. B 641, 533 (2002).

\bibitem{elvang2003}
H. Elvang and J. Polchinski, Comptes Rendus Physique {\bf 4},  405  (2003).

\bibitem{bernevig2003}
B.~A. Bernevig, J. Hu, N. Toumbas, and S.-C. Zhang, Phys. Rev. Lett. {\bf 91},
  236803  (2003).

\bibitem{hasebe2010}
K. Hasebe, Symmetry, Integrability and Geometry: Methods and Applications {\bf
  6},    (2010).

\bibitem{fabinger2002}
M. Fabinger, JHEP {\bf 2002},  037  (2002).

\bibitem{li2011}
Y. {Li} and C. {Wu}, ArXiv:1103.5422  (2011).

\bibitem{li2011a}
Y. Li, K. Intriligator, Y. Yu, C. Wu,
Phys. Rev. B {\bf 85}, 085132 (2012).


\bibitem{niemi1983}
A.~J. Niemi and G.~W. Semenoff, Phys. Rev. Lett. {\bf 51},  2077  (1983).

\bibitem{jackiw1984}
R. Jackiw, Phys. Rev. D {\bf 29},  2375  (1984).

\bibitem{redlich1984}
A.~N. Redlich, Phys. Rev. D {\bf 29},  2366  (1984).

\bibitem{semenoff1984}
G.~W. Semenoff, Phys. Rev. Lett. {\bf 53},  2449  (1984).

\bibitem{novoselov2005}
K.~S. {Novoselov} {\it et~al.}, \nat {\bf 438},  197  (2005).

\bibitem{zhang2005}
Y. {Zhang}, Y.-W. {Tan}, H.~L. {Stormer}, and P. {Kim}, \nat {\bf 438},  201
  (2005).

\bibitem{castro_neto2009}
A.~H. {Castro Neto} {\it et~al.}, Rev. Mod. Phys. {\bf 81},  109  (2009).



\bibitem{rashba1960}
E. I. Rashba, Sov. Phys. Solid State 2, 1109 (1960).

\bibitem{ast2007}
C.~R. Ast, J. Henk, A. Ernst, L. Moreschini, M.~C. Falub, D. Pacile,
P. Bruno, K. Kern, and M. Grioni, Phys. Rev. Lett. 98, 186807
(2007).

\bibitem{berg2011}
E. Berg, M. S. Rudner, and S. A. Kivelson,
Phys. Rev. B {\bf 85} 035116, (2012).

\bibitem{wu2008}
C. Wu, I. Mondragon-Shem, arXiv:0809.3532v1.

\bibitem{wu2011}
C. Wu, I. Mondragon-Shem, X. F. Zhou, Chin. Phys. Lett.
{\bf 28},  097102 (2011).



\bibitem{lin2011}
Y.-J. Lin, K. Jimenez-Garcia, I. B. Spielman, Nature 471, 83-86 (2011).


\bibitem{wang2010}
C. Wang, C. Gao, C. M.  Jian, and H. Zhai,
Phys. Rev. Lett. 105, 160403 (2010).

\bibitem{ho2011}
Tin-Lun Ho, Shizhong Zhang, Phys. Rev. Lett. 107, 150403 (2011).


\bibitem{ghosh2011}
S. K. Ghosh, J. P. Vyasanakere, and V. B. Shenoy,
Phys. Rev. A {\bf 84} 053629, (2011).


\bibitem{hu2011}
H. Hu, B. Ramachandhran, H. Pu, X. J. Liu,
Phys. Rev. Lett. {\bf 108} 010402, (2012).

\bibitem{sinha2011}
S. Sinha, R. Nath, L. Santos, arXiv:1109.2045.

\bibitem{xfzhou2011}
Xiang-Fa Zhou, Jing Zhou, Congjun Wu,
Phys. Rev. A 84, 063624 (2011).


\bibitem{xiao2010}
D. Xiao, M. C. Chang, Q. Niu,  Rev. Mod. Phys.
{\bf 82}, 1959 (2010).


\bibitem{lee2009}
D. H. Lee, Phys. Rev. Lett. 103, 196804 (2009).


\bibitem{parente2011}
V.Parente, P.Lucignano, P.Vitale, A.Tagliacozzo, F.Guinea,
Phys. Rev.B {\bf 83} 075424, (2011).

\bibitem{dalibard2011}
J. Dalibard, F. Gerbier, G. Juzeliūnas, P. Oehberg,
Rev. Mod. Phys. 83, 1523 (2011).


\bibitem{juzeliunas2011}
G. Juzeliunas, J. Ruseckas, D. L. Campbell, I. B. Spielman,
Proc. SPIE 7950, 79500M (2011).

\bibitem{ruseckas2005} J. Ruseckas, G. Juzeliunas, P. Ohberg, and M. Fleischhauer,
Phys. Rev. Lett. 95, 010404 (2005).





\end{thebibliography}
\end{document}